# THz carrier dynamics in SrTiO$_3$/LaTiO$_3$ interface two-dimensional electron gases

Ahana Bhattacharya[1], Andri Darmawan[1,2], Jeong Woo Han[1], Frederik Steinkamp[1] Nicholas S. Bingham[3,4,5], Ryan J. Suess[1], Stephan Winnerl[6], Markus E.Gruner[1,2], Eric N. Jin[5,7], Frederick J. Walker[5], Charles H. Ahn[5], Rossitza Pentcheva[1,2], and Martin Mittendorff[1,*]

[1] Universität Duisburg-Essen, Fakultät für Physik, 47057 Duisburg, Germany
[2] Universität Duisburg-Essen, Center for Nanointegration (CENIDE), 47057 Duisburg, Germany
[3] University of Maine, Department of Physics and Astronomy, Orono, ME 04469, USA
[4] University of Maine, Frontier Institute for Research in Sensor Technologies, Orono, ME 04469 USA
[5] Yale University, Department of Applied Physics, New Haven, CT 06511 USA
[6] Helmholtz-Zentrum Dresden-Rossendorf, Bautzner Landstraße 400, 01328 Dresden, Germany
[7] U.S. Naval Research Laboratory, 4555 Overlook Ave, SW Washington, D.C. 20375, USA

* email: martin.mittendorff@uni-due.de

A 2DEG forms at the interface of complex oxides like SrTiO$_3$ (STO) and LaTiO$_3$ (LTO), despite each material having a low native conductivity, as a band and a Mott insulator, respectively. The interface 2DEG hosts charge carriers with moderate charge carrier density and mobility that raised interest as a material system for applications like field-effect transistors or detectors. Of particular interest is the integration of these oxide systems in silicon technology. To this end we study the carrier dynamics in a STO/LTO/STO heterostructure epitaxially grown on Si(001) both experimentally and theoretically. Linear THz spectroscopy was performed to analyze the temperature dependent charge carrier density and mobility, which was found to be in the range of 10$^{12}$ cm$^{-2}$ and 1000 cm$^2$V$^{-1}$s$^{-1}$, respectively. Pump-probe measurements revealed a very minor optical nonlinearity caused by hot carriers with a relaxation time of several 10ps, even at low temperature. Density functional theory calculations with a Hubbard $U$ term on ultrathin STO-capped LTO films on STO(001) show an effective mass of 0.64-0.68 $m_e$.

# I Introduction

Heterostructures of complex transition metal oxides feature interface phenomena that are distinct from the bulk ranging from the formation of two-dimensional electron gases (2DEGs) to superconductivity or magnetism[1,2,3,4,5]. The 2DEGs originate from a discontinuity in the polarity of the crystal structure at the atomically sharp interfaces[6]. Depending on the composition of the oxide heterostructures, the 2DEG can host charge carriers with high mobility and high carrier density[7]. Electrostatic gating enables control over the charge carrier density, thus making it a feasible option for devices like field-effect transistors or detectors[8,9,10,11]. As it was recently shown, the 2DEG in heterostructures of the band insulator STO and the Mott insulator LTO epitaxially grown on Si(001) features a high carrier density of about $10^{12}$ cm$^{-2}$, with a mobility in the range of 100 cm$^2$V$^{-1}$s$^{-1}$ at room temperature, making it particularly interesting for devices[12]. While the dc conductivity has been studied extensively for such structures, there are relatively few studies investigating the THz conductivity[13,14]. Here we study the non-equilibrium carrier dynamics of the 2DEG at the STO/LTO interface in the vicinity of the Si(001) substrate. Linear THz time-domain spectroscopy reveals the carrier density and mobility as a function of the temperature without the need of electrical contacts, while pump-probe experiments performed in the THz spectral range give insights into the relaxation dynamics of optically excited charge carriers. For the samples investigated in this study, we found a charge carrier density on the order of $10^{12}$ cm$^{-2}$ and mobility in the range of 1000 cm$^2$V$^{-1}$s$^{-1}$, the THz conductivity is well described by the Drude model. Intraband excitation with a fluence of about 200 nJ cm$^{-2}$ leads to a rather small pump-induced increase in transmission of about 0.04%. The experimental results are complemented by density functional theory (DFT) calculations with an on-site Hubbard term to explore the electronic properties of the 2DEG. Our results show an effective mass of 0.64-0.68 $m_e$, with $m_e$ being the electron rest mass. The calculated effective mass is used as input for simulations of the experimental results via a two-temperature model, which agrees qualitatively with the experimental findings.

## II Experimental Methods and Results

The samples are grown by molecular beam epitaxy (MBE) on a low-doped and high resisitive Si substrate. The heterostructures are comprised of 4.5 unit cells (uc) STO on Si, 2 uc LTO and capped by 5 uc STO. Details of the sample growth can be found in Ref. 12, a sketch of the sample is shown in Fig. 1(a). A second sample of STO on Si, but without the intermediate 2 uc LTO layer, serves as reference for spectroscopic measurements. This way, we can exclude any contribution stemming from a 2DEG forming at the STO/Si interface or the STO/vacuum interface.

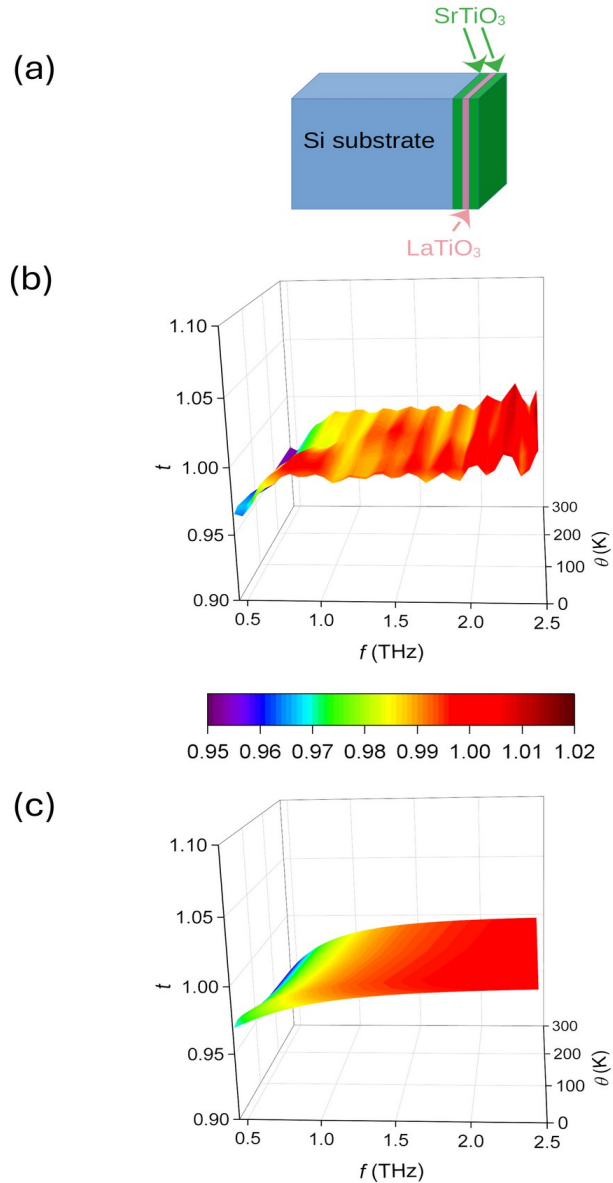

Figure 1: (a) Sketch of the sample structure. Experimental (b) and fitted (c) THz field transmission ($t$) as a function of the frequency ($f$) and temperature ($\theta$).

To characterize the charge carrier density and mobility as a function of the temperature, we performed THz time-domain spectroscopy (THz TDS). The measurements were performed in a closed-cycle cryostat equipped with z-cut quartz windows, enabling temperature dependent measurements in the temperature range from 5 K to room temperature.

The experimental transmission as a function of the frequency and the temperature is shown in Fig. 1(b), the oscillations of the transmission, appearing as ripples along the temperature axis, are caused by multiple reflections within the sample. Even though 2DEGs in STO are characterized by a rather complex interplay of electron-electron and electron-phonon scattering[15], the THz transmission can be well described with a simple Drude model. To extract the carrier density and mobility from the experimental results, we fit a Drude conductivity in combination with a thin film model to the experimental results via[16]

$$t(\omega) = \left| \frac{E_{sample}(\omega)}{E_{reference}(\omega)} \right| = \left| \frac{n_{Si}+1}{n_{Si}+1+Z_0 \sigma(\omega)} \right|, \qquad \text{Eq. 1}$$

where $t(\omega)$ is the THz field transmission obtained from the sample and reference spectra $E_{sample}(\omega)$ and $E_{reference}(\omega)$, respectively. The refractive index of the substrate is represented by $n_{Si}$, $Z_0$ is the free-space impedance and $\sigma(\omega)$ the sheet conductivity of the 2DEG. The latter is derived from the Drude model via

$$\sigma(\omega) = \frac{n e^2 \tau}{m^*(1-i\omega\tau)}, \qquad \text{Eq. 2}$$

where $m^*$ represents the effective mass, $\tau$ the momentum scattering time, $n$ is the sheet carrier density and $e$ the electric charge of an electron. The effective mass $m^*$ is derived from the DFT simulations as described below. It is obtained by averaging over the calculated values of effective masses of the 2DEG along Γ-M and Γ-X directions (the band structure is depicted in Fig. 4c). As a result, we get the charge carrier density and mobility as a function of the temperature as shown in Fig. 2(a) and (b), respectively.

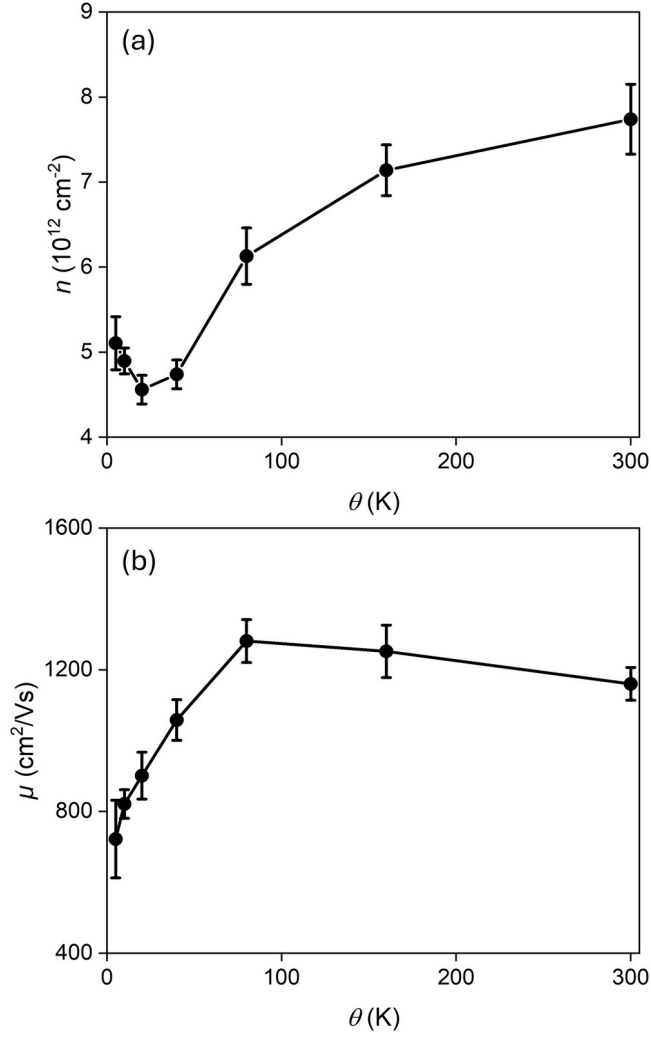

Figure 2: Extracted (a) charge carrier density ($n$) and (b) mobility ($\mu$) as a function of the temperature ($\theta$).

The mobility is observed to increase moderately with temperature and reaches significantly above 1000 cm²V⁻¹s⁻¹ before decreasing to about 1160 cm²V⁻¹s⁻¹ at room temperature. The charge carrier density decreases slightly with temperature on heating the sample from 5K to 20K. On heating beyond 20K, charge carrier density increases with temperature, reaching a value of approximately $7.7 \cdot 10^{12}$ cm⁻² at room temperature, which is consistent with earlier findings[12]. The magnitude of the mobility and carrier density lies between those found for the two-carrier model (mobility of 10000 in the silicon and 100 in the oxide layers) used in Ref. 12.

To measure the non-equilibrium carrier dynamics in the 2DEG, we performed pump-probe experiments at the free-electron laser (FEL) facility FELBE at Helmholtz-Zentrum Dresden-Rossendorf. To efficiently excite the 2DEG via free-carrier absorption and avoid heating via phonon absorption[17], we tuned the FEL to 1.35 THz, corresponding to a photon energy of about 5.6 meV. The FEL provides a continuous pulse train with a repetition rate of 13 MHz[18], a small fraction of about 2% of the FEL power is split off to serve as probe, the polarization is rotated by 90° after passing a delay stage. The majority of the FEL power is guided to a parabolic mirror, focusing both, pump and probe beam, on the sample. While the pump beam is blocked behind the sample, the probe beam is guided through an additional polarizer in order to minimize scattered pump radiation, before detection with a bolometer. The sample is mounted in a flow cryostat to maintain a sample temperature of 10 K. In order to avoid spurious pump-probe signals stemming from a potential 2DEG at the STO/ Si(001) interface and/or at the STO/vacuum interface, we performed measurements on the reference sample, observing no measurable change in transmission.

The experiments are performed at four different pump fluences of 16 nJ cm$^{-2}$, 53 nJ cm$^{-2}$, 106 nJ cm$^{-2}$, and 185 nJ cm$^{-2}$, the corresponding pump-probe signals are shown in Fig. 3(a). As can be seen, only a minor pump-induced change in transmission of about 0.04% is observed at the highest pump fluence. To measure these very small pump-induced changes in transmission, we averaged over 40 measurements at each fluence. After the peak of the pump-probe signal is reached, the signal decays on a time scale of about 50 ps, indicating a fast cooling of the hot carriers. The lines in Fig. 3(a) serve as guide to the eye, indicating standard pump-probe signals that can be described by an error function for the rising edge and an exponential decay for the carrier relaxation. Due to the poor signal-to-noise ratio, the carrier relaxation time cannot be determined precisely. The maximum of the pump-probe signal scales with the square-root of the pump fluence, which is represented by a phenomenological fit as red solid line in Fig. 3(b).

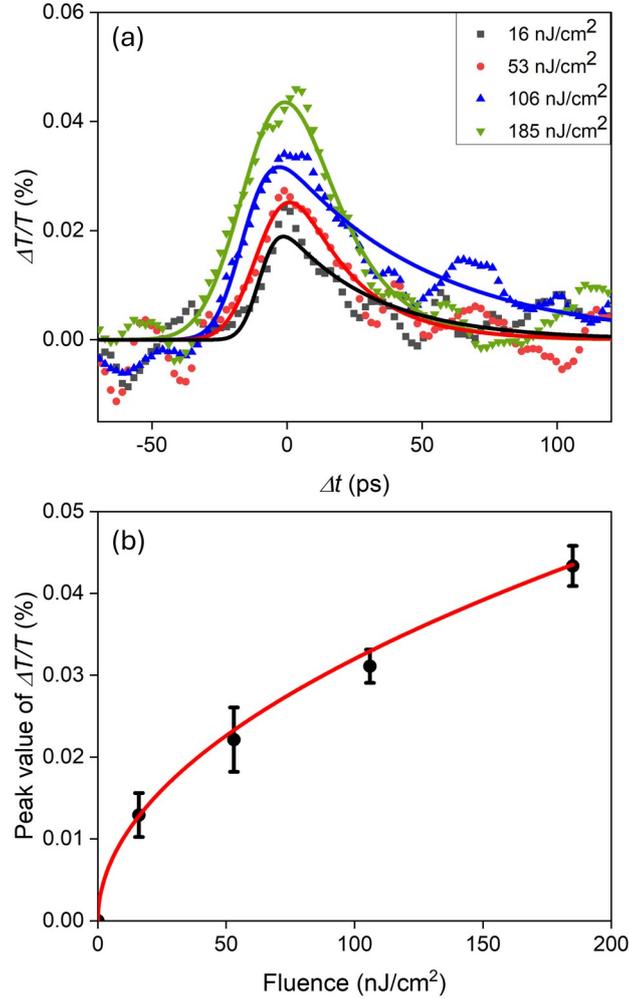

Figure 3: Pump-induced change in transmission ($\Delta T/T$) as a function of the delay time ($\Delta t$) for various pump fluences. The solid lines serve as guides to the eye. (b) Maximum pump-induced change in transmission as a function of the applied pump fluence, the red line represents a phenomenological square root fit.

III Theoretical Results and Discussion

Density functional theory calculations were performed on STO/LTO/STO/Si(001) as well as STO/Si(001) using the Vienna ab-initio package (VASP)[19,20,21] which implements the projector augmented wave (PAW)[22,23] method and pseudopotentials. The PBE[24] exchange-correlation functional within the generalized gradient approximation was used with an on-site Hubbard $U$ term in the rotationally invariant formulation of Dudarev et al.[25]. Effective corrections $U_{eff}$= 5 eV and 8 eV are applied on the Ti $3d$ and La *4f* states, respectively, consistent with previous work[26,27]. We employed a *p*(2×2) lateral unit cell consisting of four inequivalent Ti sites per layer in order to allow octahedral tilts and rotations, as well as an antiferromagnetic G-type ordering for the bulk LTO phase. The lateral lattice constant is fixed to the experimental lattice constant of Si (5.43Å) exposing the STO and LTO films to -1.70% and -4.04% compressive strain, respectively. Asymmetric slabs were utilized with a 1ML STO capping layer, followed by 2MLs of LTO and 3MLs of STO on 9 MLs reconstructed Si(001) substrate, whose bottom side is passivated by H. A vacuum region of 30Å is added to avoid interactions between the slab and its periodic images. Additionally, we have considered 3ML STO/Si(001) as a reference system. We model the interface between STO and Si(001)[28] by a SrO-termination at the reconstructed Si(001) interface, as previously reported by Chen et al[29]. Overall, the studied systems contain 164 and 104 atoms for STO/LTO/STO/Si(001) and STO/Si(001), respectively. We used an energy cut-off of 500 eV and sampled the Brillouin zone using a 5×5×1 Γ-centered *k*-mesh. The ionic positions were fully relaxed.

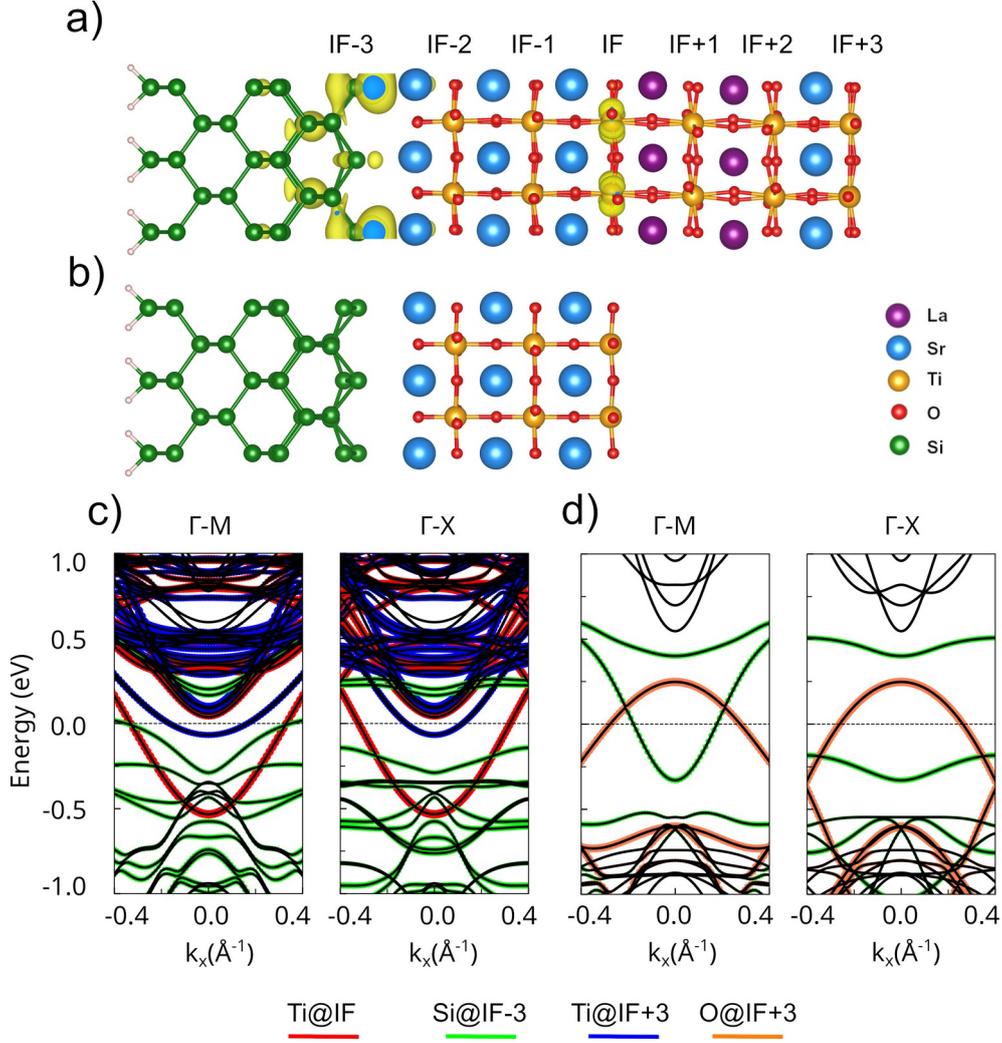

Figure 4: Spin density of (a) STO/LTO/STO on reconstructed Si(001) and (b) STO on reconstructed Si(001), (integrated between -0.6 eV and the Fermi energy at $E_F=0$ with isosurface value of 0.0006 e/Å$^3$). The lower panels show the site-resolved band structure along the Γ-M and Γ-X directions in (c) STO/LTO/STO/Si(001) and (d) STO/Si(001).

Figs. 4a and b displays side views of the relaxed systems. For STO/LTO/STO/Si(001) the spin-density, integrated from -0.6 eV to $E_F$, displays a significant contribution at the LTO/STO interface with $d_{xy}$ orbital polarization at the Ti sites as well as at the topmost Si layers. The band structure shown in Fig.4c indicates that several conduction bands cross the Fermi level and thus contribute to the metallicity of the system and the formation of a 2DEG. The main contribution arises from Ti $3d$ states at the LTO/STO interface with some participation of Si at the STO/Si(001) interface (IF-3). The carriers at the LTO/STO interface have mainly $d_{xy}$ character, wheereas the $d_{xz}$ and $d_{yz}$ orbitals are lying above the Fermi level, while for LTO/STO(001) without a Si substrate, the carriers involved in the formation of the 2DEG are predominantly $d_{xy}$ and $d_{xz+yz}$ orbitals[30].

Furthermore we have calculated the effective mass of the bands at the LTO/STO interface contributing to the 2DEG along the Γ-M and Γ-X directions- from the band stucture depicted in Fig. 4c. The calculated effective mass along Γ-M direction is $0.64 m_e$ while along Γ-X direction is $0.68 m_e$. Those bands are stemming from the $3d_{xy}$ orbitals of Ti at the LTO/STO interface. In contrast, the reference system shows only two bands crossing at $E_F$, a hole band contributed by O at the surface (IF+3) as well as an electron band of Si at the interface (IF-3) as seen in Fig. 4d. Our results show that the 2DEG at the LTO/STO interface has predominantly Ti $d_{xy}$ character.

To simulate the dynamics of the 2DEG, we perform calculations based on a two-temperature model via $\alpha_e \theta \frac{d\theta}{dt} + \beta(\theta - \theta_l) = A(\theta) I(t)$,            Eq. 3 where $\alpha_e$ represents the specific heat of the 2DEG, which can be derived from the effective mass, assuming a parabolic dispersion relation. The temperature of the 2DEG and the lattice are represented by $\theta$ and $\theta_l$, respectively. $\beta = 1.4 \cdot 10^4 \, W \, m^{-2} \, K^{-1}$ is a fit parameter for the carrier relaxation time, $A(\theta)$ is the absorption of the 2DEG derived from the thin film model at the photon frequency applied in the experiment (1.35 THz), and $I(t)$ describes the temporal evolution of the pump pulse. The temperature dependent parameters for the absorption, i.e. the carrier density and mobility, are interpolated from the THz TDS results shown in Fig. 2. From the temporal evolution of the electron temperature, we calculated the pump-induced change in transmission via the thin-film model shown in Fig. 5(a), Fig. 5(b) shows the maximum change in transmission as a function of the pump fluence. While the theoretical results qualitatively reproduce the experimental results including the fluence dependence, the absolute values are about five times higher than experimentally observed. From the specific heat mentioned above, we estimate a maximum increase of the electron temperature of about 18 K. We note that the quantitative deviation between the two-temperature model and the experimental results is likely related to the fundamental difference between the pump-probe experiments and the THz TDS measurements: while the THz TDS measurements were taken at equilibrium, i.e. both, electrons and lattice have the same temperature, the pump-probe measurements probe a hot carrier distribution in a cold lattice.

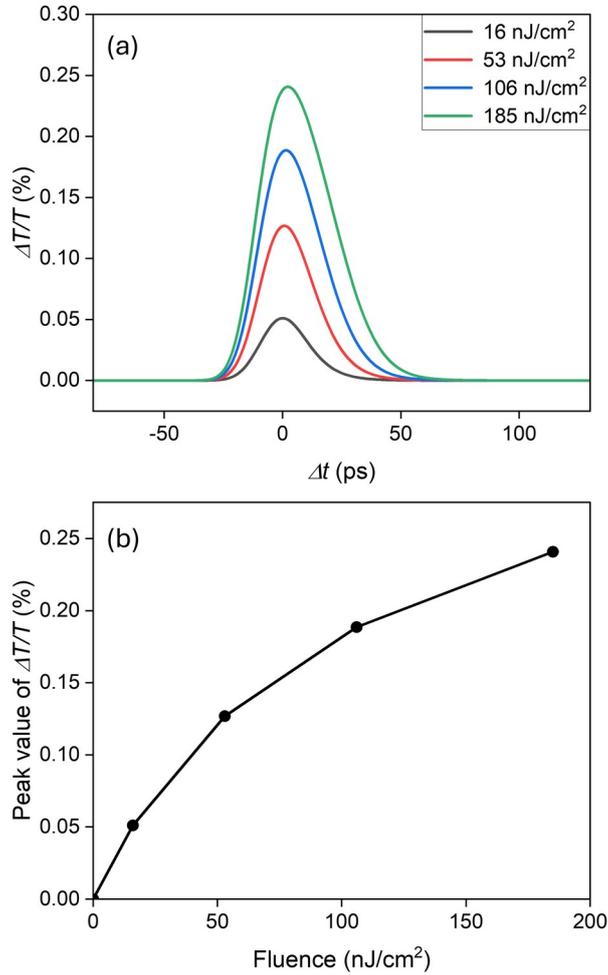

Figure 5: (a) Simulated pump-induced change in transmission ($\Delta T/T$) derived from a two-temperature model. (b) Simulated maximum of the pump-induced change in transmission as a function of the applied pump fluence.

The pump-induced change in transmission is rather small compared to other 2DEGs based on conventional heterostructures[31,32]. We attribute the remarkably small pump-probe signals to the temperature dependence of the oxide-interface 2DEG in our STO/LTO heterostructure. In the interfacial 2DEGs studied here, we see the carrier density initially decreases with increasing temperature at first and reaches a minimum value at 20 K. Only for temperatures above this minimum value is an increase in carrier density observed (cf. Fig. 2(a)). The mobility in the STO/LTO 2DEG increases until a temperature of about 80 K beyond which it shows a small drop, indicating weak electron-phonon scattering. The decrease in mobility below 80K as well as the initial decrease of carrier density with temperature point towards the presence of ionized impurities at the interface. The trends observed in the carrier density and mobility in the STO/LTO 2DEG are counteracting on the transmission/absorption of THz radiation, resulting in the very small pump-induced change in transmission observed in our pump-probe experiments.

## IV. Summary

In conclusion, we investigated the physical properties of the 2DEG formed at the STO/LTO interface via THz TDS and high intensity THz pump-probe measurements. The experiments are complemented by DFT+U calculations of the 2DEG that find an effective mass of 0.64-0.68 $m_e$. The optical properties and the dynamical evolution of the 2DEG are modelled via thin-film model and two-temperature model, respectively. We found a surprisingly low impact of the intraband excitation on the sample conductivity, that resulted in small pump-induced changes of the transmission that were less than 0.05%.

Acknowledgment

This study was funded by the Deutsche Forschungsgemeinschaft (DFG, German Research Foundation)—Project-ID 278162697—SFB1242. We thank J. Michael Klopf and the ELBE team for their assistance. Work at Yale supported by the Office of Naval Research Multidisciplinary University Research Initiative to support the EXtreme Electron DEvices (EXEDE) program (synthesis and characterization) and by NSF DMR-2412358 (analysis).